\documentclass[12pt]{article}

\evensidemargin \oddsidemargin
\usepackage[T1]{fontenc} 
\usepackage{amsmath}
\usepackage{amssymb}
\usepackage{hyperref}
\usepackage{latexsym}
\usepackage{graphicx}
\usepackage{color}
\usepackage{braket}
\usepackage{amssymb}
\usepackage{bbold}

\newcommand{\Imag}{\mathop{\mathrm{Im}}}

\newcommand{\mL}{\mathcal{L}}

\newcommand{\mM}{\mathcal{M}}

\newcommand{\mO}{\mathcal{O}}
\newcommand{\mP}{\mathcal{P}}

\begin{document}
\thispagestyle{empty}

\def\thefootnote{\fnsymbol{footnote}}

\vspace{0.5cm}

\begin{center}
{\large\sc {\bf  
Coupling $WW$, $ZZ$ unitarized amplitudes to $\gamma\gamma$ in the TeV region
}}

\vspace{1cm}

{\sc
Rafael L. Delgado, Antonio Dobado and Felipe J. Llanes-Estrada
}

\vspace*{.7cm}

{\sl
Departamento de F\'isica Te\'orica I\\
Universidad Complutense de Madrid,\\
 Plaza de las Ciencias 1,
28040 Madrid, Spain

}
\end{center}

\vspace*{0.1cm}

\begin{abstract}
We define and calculate helicity partial-wave amplitudes for processes linking the Electroweak Symmetry Breaking Sector (EWSBS) to $\gamma\gamma$, employing (to NLO) the Higgs-EFT (HEFT) extension of the Standard Model and the Equivalence Theorem, while neglecting all particle masses. The resulting amplitudes can be useful in the energy regime ($500\,{\rm GeV}-3\,{\rm TeV}$). We also deal with their unitarization so that resonances of the EWSBS can simultaneously be described in the $\gamma\gamma$ initial or final states. Our resulting amplitudes satisfy unitarity, perturbatively in $\alpha$, but for all $s$ values. In this way we improve on the HEFT that fails as interactions become stronger with growing $s$ and provide a natural framework for the decay of dynamically generated resonances  into $WW$, $ZZ$ and $\gamma\gamma$ pairs.
\end{abstract}

\def\thefootnote{\arabic{footnote}}
\setcounter{page}{0}
\setcounter{footnote}{0}

\newpage

\section{Introduction}
\subsection{The Electroweak Symmetry Breaking Sector}
Electroweak symmetry breaking happens at a scale of $v=246$ GeV for reasons still unsettled, and the LHC is trying to discern whether the Higgslike scalar boson found 
there~\cite{ATLAS,CMS,Aad:2013wqa,Chatrchyan:2013lba}
 couples to other known particles as dictated by the Standard Model.  If the LHC finds new particles or perhaps broad resonances in the TeV region under exploration, it is natural (by their energy scale) to guess that they play a role in breaking electroweak symmetry. 

Meanwhile, it makes sense to formulate theory in terms of the particles already known to be active in that Electroweak Symmetry Breaking Sector, namely the new Higgslike boson $h$ and the longitudinal components of the $W$ and $Z$ electroweak bosons. The resulting, most general, effective field theory that does not assume $h$ to be part of an electroweak doublet, has come to be known as Higgs Effective Field Theory (HEFT)~\cite{Alonso:2015fsp,Gavela:2014uta,Pich:2013fba,Buchalla:2013rka,Buchalla:2015qju,YellowReport}, and has been built upon the old Higgsless Electroweak Chiral Lagrangian~\cite{Appelquist}. An effort to constrain the parameters of this HEFT from low-energy observables is underway employing LHC data~\cite{Brivio:2016fzo}. We adopt the parity-conserving HEFT as our starting point.

To reduce the complexity of computations and since we are not aiming at a precision description of $WW/hh$ threshold observables, but rather at the possible resonance region above 500 GeV, we adopt the Equivalence Theorem~\cite{ET}. This is valid for $s\gg M_h^2,M_W^2,M_Z^2\simeq (100\,{\rm GeV})^2$, and allows identifying the longitudinal gauge bosons with the pseudo-Goldstone bosons of symmetry breaking $\omega^a$ ($a=1\dots 3$) in their scattering amplitude $T$. For example one has:
\begin{equation}\label{eq:EquivalenceTheorem}
  T(W^i_L W^j_L\to W_L^k W_L^l) = T(\omega^i\omega^j\to\omega^k\omega^l) + \mO\left(\frac{M_W}{\sqrt{s}}\right)\ .
\end{equation}

In our recent work analysing the EWSBS~\cite{Delgado:2013hxa,Delgado:2013loa,Delgado:2015kxa} we established that, for any parameter choice separating from the Standard Model, the theory becomes strongly interacting at sufficiently high energy, and resonances may appear. Given the mass gap between the visible EWSBS and those resonances, it makes sense to restrict ourselves to Higgs constant self-couplings that count as order $M_h^2$, and are thus negligible for $s\gg M_h^2$, so that the Higgs also couples derivatively in our energy interval (as in Composite Higgs or dilaton models). Apart from this assumption our discussion remains general. The Lagrangian will be exposed below in section~\ref{chLagr}.

In principle, HEFT is a usable theory through $E\sim 4\pi v =3$ TeV, but if new strong interactions and resonances appear in that interval, its applicability region quickly contracts. Even at low energies, truncated HEFT may run into convergence difficulties. These are not alleviated much by increasing the order of the chiral expansion, and on the contrary the number of chiral parameters swiftly increases reducing predictive power. We therefore follow a different strategy to extend the low energy regime by using dispersion relations (DR) compatible with analyticity and unitarity. This approach is  extremely useful in the original hadron ChPT~\cite{Truong:1988zp} and we have long advocated it for the SBS of the strongly interaction sector of the SM~\cite{Dobado:1989gr}. 

Dispersion relations are identities that do not include all dynamical information, but they become much more predictive when the low-energy scattering amplitudes are known (for example from the HEFT). Even so, some model dependence remains in the treatment of amplitude cuts outside the physical $s$ (left cut, LC). To constrain it, we employ two different methods, the Inverse Amplitude Method (IAM) and the N/D method. We have detailed both of them in this context in~\cite{Delgado:2015kxa} where complete discussion and references may be found, as well as further unitarization strategies. As will also be exposed later (see figure~\ref{fig:comparemethods}), both methods are qualitatively equivalent, but  to describe the same resonance they need parameters of the underlying NLO HEFT that are different from each other by 25\%.  

In the end, the resulting unitarized HEFT (UHEFT) provides an analytical and unitary description of higher energy dynamics which is essentially unique up to the first resonances. These appear as poles in the second Riemann sheet thanks to the adequate analytical behavior of the amplitudes.

\subsection{Coupling to $\gamma\gamma$}

The $\gamma\gamma$ channel is by itself not part of the EWSBS, but because photons propagate to the detectors (being reconstructed, e.g. at the electromagnetic calorimeters) they are direct messengers from the collision in the final state.  Studies of new particles decaying into two photons have been pursued since the dawn of particle physics~\cite{Yang:1950rg}.

Conversely, photons also provide interesting production mechanisms from the initial state. With slight virtuality they accompany high-energy beam particles: the photon can be thought of as a parton of the proton~\cite{Manohar:2016nzj} or the electron in $pp$ and $e^-e^+$ colliders, respectively. Thus, high-energy colliders can, in a sense, be thought of as photon colliders. The small electromagnetic $\alpha$ lowers the photon flux, but in exchange the initial state is very clean and directly couples to the EWSBS (since the $W^\pm$ are charged particles). 
For example, the CMS collaboration~\cite{Khachatryan:2016mud} is already setting bounds to anomalous quartic gauge couplings from an analysis of precisely $\gamma\gamma \to W^+ W^-$.
Moreover, thanks to Compton backscattering, photon colliders driven by lepton beams are perhaps also a future option~\cite{Telnov:2016lzw,Gronberg:2014yfa}.

Thus, their coupling to the EWSBS is of much interest. Within the context of the HEFT, the perturbative Feynman amplitudes at the one-loop level have already been reported in~\cite{Delgado:2014jda}.

In this work we extend the amplitudes to the resonance region. Because unitarity is most easily expressed in terms of partial waves, and because the partial-wave series converges quickly in the ``low-energy'' domain where HEFT is valid, we have projected the EWSBS as well as the $\gamma\gamma$ over good angular momentum $J$. In the case of the Goldstone or the Higgs bosons, $L=J$, but when photons are involved, their intrinsic spin is also at play. We have employed the helicity basis to carry out the computations.

While custodial isospin is presumably conserved by the EWSBS (as suggested by LEP), the electromagnetic coupling to the $\gamma\gamma$ state violates its conservation. Still, we can label the partial wave amplitudes from the initial $\omega\omega$-state isospin in photon-photon production (or the final $\omega\omega$ at a photon collider).

The helicity basis and the corresponding amplitudes are constructed below in section~\ref{sec:gammagamma}. Their partial wave projections in turn appear in section~\ref{sec:PartialWaves:gammagamma}. We show their single- and coupled-channel unitarization in section~\ref{sec:unitarity} and some selected numerical examples thereof in section~\ref{sec:numerics}; at last, we add a few remarks in section~\ref{sec:conclusions}.

\section{The chiral Lagrangian and its parameterizations}\label{chLagr}
First we quote from Ref.~\cite{Delgado:2013loa} 
the effective Lagrangian describing the low-energy dynamics of the four light modes: three would-be Goldstone Bosons $\omega^a$ (WBGBs) and the Higgs-like particle $h$. This particle content is valid for the energy range $M_h,M_W,M_Z\simeq (100\,{\rm GeV})^2 \ll s\ll 4\pi v\simeq 3\,{\rm TeV}$ and exhausts the known Electroweak Symmetry Breaking Sector. Resonances of these particles' scattering are possible in this interval and we will describe them employing scattering theory fundamented on the effective Lagrangian instead of introducing them as new fields. The starting point to expose the Lagrangian for the $\omega$ and $h$ bosons, whose elements are immediately discussed, may be taken as
\begin{equation}\label{chLagr:effective:WBGBs}
 \mL = \frac{v^2}{4}\mathcal{F}(h/v) {\rm Tr}[(D_\mu U)^\dagger D^\mu U] + 
                \frac{1}{2}\partial_\mu h\partial^\mu h - V(h),
\end{equation}
where the vacuum constant is $v=246\,{\rm GeV}$, and the arbitrary function 
\begin{equation} \label{thefunc_g}
\mathcal{F}(h/v)=1+2a\frac{h}{v}+b\left(\frac{h}{v}\right)^2+\dots\ 
\end{equation}
is analytic around vanishing scalar field. The NLO computation for the WBGB sector is quoted in Refs.~\cite{Delgado:2013hxa,Delgado:2015kxa}. Note the usage of the spherical parameterization\footnote{In Ref.~\cite{Delgado:2014jda}, we also employed the exponential parametrization of the coset for the $\gamma\gamma$ scattering. While intermediate results (i.e., the Feynman diagrams) are different, the on-shell amplitudes are exactly the same for both parametrizations.}. The extension that includes $\gamma\gamma$ states can be found in Ref.~\cite{Delgado:2014jda}, the covariant derivative being
\begin{equation}\label{chLagr:DU}
  D_\mu U = \frac{i\partial_\mu\omega_i\tau^i}{v}+i\frac{g}{2}W_{\mu,i}\tau^i - i\frac{g'}{2} B_\mu\tau^3\cdots  \ .
\end{equation}
Here,  the dots represent terms of higher order in $(\omega^a/v)$ and whose precise form 
depends on the particular parametrization of $U$.
At last, we note the definition of charge basis, $\omega^\pm = \frac{\omega^1\mp i\omega^2}{\sqrt{2}}$, $\omega^0 = \omega^3$.  Thus we are using  a $SU(2)_L \times U(1)_Y$ gauged non-linear sigma model corresponding to the coset $SU(2)_L \times SU(2)_R/ SU(2)_{L+R}$ coupled to the $h$ singlet, where $SU(2)_{L+R}$ is called the custodial group.

Different values of the parameters $a$ and $b$ in Eq.~(\ref{thefunc_g}) make the Lagrangian density in Eq.~(\ref{chLagr:effective:WBGBs}) represent the low-energy limit of different theoretical models (and the NLO parameters specified shortly will depend on the underlying theory). For instance, $a^2=b=0$ corresponds to the old, Higgsless Electroweak Chiral Lagrangian~\cite{Appelquist} (that had no explicit Higgs field and thus seems ruled out), $a^2=1-(v/f)^2$, $b=1-2(v/f)^2$ is the low-energy limit of a $SO(5)/SO(4)$ Minimal Composite Higgs Model~\cite{SO(5)}, $a^2=b=(v/f)^2$ can be obtained from dilaton-type models~\cite{Grinstein}, and finally $a^2=b=1$ represents the SM with a light Higgs (current experimental situation).

There is no strong direct limit over the $b$ parameter, because of the difficulty of measuring a 2-Higgs state. However, an indirect limit arises because of the coupling between the $hh$ decay and the elastic $\omega\omega$ scattering, as we showed in earlier work~\cite{Delgado:2014dxa}. The current direct claimed limits over the $a$ parameter, at a confidence level of $2\sigma$ ($\approx 95\%$) are, from CMS~\cite{Khachatryan:2014jba}, $a\in (0.87,\, 1.14)$; and from ATLAS~\cite{ATLAS:2014yka}, $a\in (0.96,\,1.34)$. Actual experimental analysis may be tracked from~\cite{Giardino:2013wva}, that also details LHC constraints over a number of SM extensions.

\subsection{WBGBs scattering and coupling to $\gamma\gamma$}\label{subsec:couplegammagamma}

The one-loop computation for $\omega\omega\to\omega\omega$, $\omega\omega\to hh$ and $hh\to hh$ processes was reported in~\cite{Delgado:2013hxa,Delgado:2015kxa}. Because of the Equivalence Theorem regime, $e^2,g^2,g^{'2}\ll s/v^2$, the electric charge coupling the photon can be 
introduced as a perturbation.
Thus, the strong physics of the $\omega\omega$ (longitudinal $W_L$ modes) and $hh$ sector dominates over the transverse modes ($W_T$, $\gamma$) and provides the driving force to saturate unitarity. One can then work to leading non-vanishing order when incorporating the transverse modes. The minimum set of counterterms needed to renormalize those scattering amplitudes to one loop is that corresponding to the  $a_4$, $a_5$, $g$~\footnote{Not to be confused with the $SU(2)_L$ gauge coupling.}, $d$ and $e$ parameters (see Refs.~\cite{Delgado:2013hxa,Delgado:2015kxa}).

On Ref.~\cite{Delgado:2014jda} we extended the effective NLO Lagrangian for the Higgs and WBGBs~\cite{Delgado:2013hxa,Delgado:2015kxa}  
{by including transverse gauge bosons to account for the $\gamma\gamma\to zz$ and $\gamma\gamma\to\omega^+\omega^-$ processes.  Concentrating only on $\gamma\gamma$ 
the effective Lagrangian becomes
\begin{equation}\label{chLagr:Lagr:WBGB:gamma:spherical:param}
\begin{split}
\mL_2(\omega,h,\gamma) ={} & %
       \frac{1}{2}\partial_\mu h\partial^\mu h
      +\frac{1}{2}\mathcal{F}(h/v)(2\partial_\mu\omega^+\partial^\mu\omega^-
      +\partial_\mu\omega^0\partial^\mu\omega^0) \\
     &+\frac{1}{2v^2}\mathcal{F}(h/v)(\partial_\mu\omega^+\omega^- 
      +\omega^+\partial_\mu\omega^- 
      +\omega^0\partial_\mu\omega^0)^2 \\
     &+ie \mathcal{F}(h/v)A^\mu (\partial_\mu\omega^+\omega^- - \omega^+\partial_\mu\omega^-)
      +e^2 \mathcal{F}(h/v)A_\mu A^\mu\omega^+\omega^-,
\end{split}
\end{equation}
where the photon field is given by $A_{\mu}= \sin \theta_W W_{\mu,3}+ \cos \theta_W B_{\mu}$ with $\sin \theta_W = g' / \sqrt{g^2+ g'^2}$. If the Lagrangian of Eq.~(\ref{chLagr:Lagr:WBGB:gamma:spherical:param}) is employed at NLO, a counterterm NLO Lagrangian is in principle necessary to guarantee the order by order renormalizability, as customary in EFT. This brings in the additional $a_1$, $a_2$, $a_3$ and $c_\gamma$ counterterms (see Ref.~\cite{Delgado:2014jda}),
\begin{multline}
\mL_4 = %
  a_1 {\rm Tr}(U \hat{B}_{\mu\nu} U^\dagger \hat{W}^{\mu\nu})
  + i a_2 {\rm Tr} (U \hat{B}_{\mu\nu} U^\dagger [V^\mu, V^\nu ]) 
  - i a_3  {\rm Tr} (\hat{W}_{\mu\nu}[V^\mu, V^\nu]) \\
 -\frac{c_{\gamma}}{2}\frac{h}{v}e^2 A_{\mu\nu} A^{\mu\nu} + \dots,
\label{eq.L4}
\end{multline}
where:
\begin{eqnarray}
\hat{W}_{\mu\nu} &=& \partial_\mu \hat{W}_\nu - \partial_\nu \hat{W}_\mu
 + i  [\hat{W}_\mu,\hat{W}_\nu ],\;\hat{B}_{\mu\nu} = \partial_\mu \hat{B}_\nu -\partial_\nu \hat{B}_\mu ,\label{fieldstrength}\\
\hat{W}_\mu &=& g \vec{W}_\mu \vec{\tau}/2 ,\;\hat{B}_\mu = g'\, B_\mu \tau^3/2 ,
\label{EWfields}\\
V_\mu &=& (D_\mu U) U^\dagger ,\;\label{VmuandT} \\
A_{\mu\nu} &=& \partial_\mu A_\nu - \partial_\nu A_\mu\ .
\end{eqnarray}
Eq.~(\ref{eq.L4}) can be expanded as
\begin{multline}
\mL_4 = %
\frac{e^2a_1}{2v^2}A_{\mu\nu}A^{\mu\nu}\left(v^2 - 4\omega^+\omega^-\right) %
 + \frac{2e(a_2-a_3)}{v^2}A_{\mu\nu}\left[%
         i\left(\partial^\nu\omega^+\partial^\mu\omega^- - \partial^\mu\omega^+\partial^\nu\omega^- \right) %
   \right. \\ \left. %
        +eA^\mu\left( \omega^+\partial^\nu\omega^- + \omega^-\partial^\nu\omega^+ \right)
        -eA^\nu\left( \omega^+\partial^\mu\omega^- + \omega^-\partial^\mu\omega^+ \right)
        \right] -\frac{c_{\gamma}}{2}\frac{h}{v}e^2 A_{\mu\nu} A^{\mu\nu} .
\end{multline}
The chiral counting for the EFT yielding $\omega\omega\to \gamma\gamma$ is compared to that for the elastic $\omega\omega \to \omega\omega$ process in figure~\ref{fig:counting}.

\begin{figure}
\centerline{
\includegraphics[width=6cm]{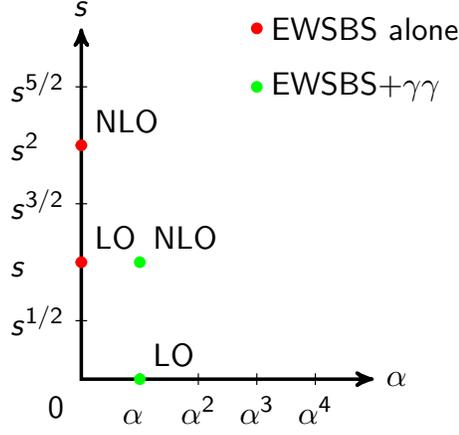}}
\caption{\label{fig:counting} A sensible counting for $\omega\omega\to \gamma \gamma$, in the energy region around $\mO(0.5\,{\rm TeV})$ of interest for the LHC, is to take $\alpha_{EM}$ and $s$ as small quantities with $\alpha_{EM}$ (of smaller size)  only to first order. The resulting counting (green dots) is compared to that for the purely Goldstone boson processes $\omega\omega\to \omega\omega$ (red dots).}
\end{figure}

Current ($2\sigma$) bounds on those NLO parameters are $c_\gamma\in\left(\frac{-1}{16\pi^2},\frac{0.5}{16\pi^2}\right)$~\cite{Buchalla:2015qju}; $a_1<10^{-3}$, $a_2\in(-0.26,0.26)$ and $a_3\in(-0.1,0.04)$~\cite{Fabbrichesi:2015hsa}. It is practical to quote these bounds for the $a_i$ in terms of the only combination that will be needed in this work which is (conservatively adding them) $(a_1-a_2+a_3)\in (-0.36,0.3)$. We will employ these limits when we illustrate the parameter dependence later on in section~\ref{sec:numerics}.

\section{Matrix elements for $\gamma\gamma$ to $\omega\omega$ and $hh$ scattering at NLO}\label{sec:gammagamma}
The one-loop perturbative amplitudes for the production of $\gamma\gamma$ from the EWSBS can be read off (by time reversal invariance) from those for $\gamma\gamma\to \omega\omega$ scattering, computed in~\cite{Delgado:2014jda}. Both $\gamma\gamma\to zz$ and $\gamma\gamma\to\omega^+\omega^-$ amplitudes were there decomposed as
\begin{equation}\label{A:scatter:gammagamma:lorentz}
  \mM_{\lambda_1\lambda_2} = iT = ie^2\left(\epsilon_1^\mu\epsilon_2^\nu T_{\mu\nu}^{(1)}\right) A %
       +ie^2\left(\epsilon_1^\mu\epsilon_2^\nu T_{\mu\nu}^{(2)}\right) B,
\end{equation}
with Lorentz structures
\begin{subequations}
\begin{align}
  \left(\epsilon_1^\mu\epsilon_2^\nu T_{\mu\nu}^{(1)}\right) &= %
     \frac{s}{2}(\epsilon_1\epsilon_2) - (\epsilon_1 k_2)(\epsilon_2 k_1) %
\label{A:scatter:gammagamma:LorentzStructure:1}\\
  \left(\epsilon_1^\mu\epsilon_2^\nu T_{\mu\nu}^{(2)}\right) &= %
     2s(\epsilon_1\Delta)(\epsilon_2\Delta)-(t-u)^2(\epsilon_1\epsilon_2) %
   -2(t-u)[(\epsilon_1\Delta)(\epsilon_2 k_1)-(\epsilon_1 k_2)(\epsilon_2\Delta)] . %
\label{A:scatter:gammagamma:LorentzStructure:2}
\end{align}
\end{subequations}
Here, $e=\sqrt{\alpha/4\pi}\approx 0.303$ is the electric charge; $\epsilon_i(\lambda_i)$, $\lambda_i= \pm 1$ and $k_i$ are the polarization state vector, the helicity and the 4-momentum of each photon with $i=1,\,2$; $p_i$, the 4-momenta of the Gauge boson ($i=1,\,2$); and $\Delta^\mu = p_1^\mu - p_2^\mu$.

The $\gamma\gamma\to zz$ process, at order $\mO (e^2)$ and leading chiral order, vanishes because the $Z$ is a neutral particle,
\begin{equation}\label{A:scatter:gammagamma:to:zz:Atree}
  \mM (\gamma\gamma\to zz)_{\rm LO} = 0 .
\end{equation}
The NLO contribution depends on $c_\gamma$,
\begin{subequations}\label{A:scatter:gammagamma:to:zz:Aloop}
\begin{align}
  A(\gamma\gamma\to zz)_{\rm NLO} &= %
      \frac{2ac_\gamma^r}{v^2} + \frac{a^2-1}{4\pi^2 v^2} \equiv A_N 
\label{A:scatter:gammagamma:to:zz:Aloop:A}\\
  B(\gamma\gamma\to zz)_{\rm NLO} &= 0 .
\label{A:scatter:gammagamma:to:zz:Aloop:B}
\end{align}
\end{subequations}

For $\gamma\gamma\to\omega^+\omega^-$ (the only other process allowed by charge conservation), at order $\mO(e^2)$,
\begin{equation}\label{A:scatter:gammagamma:to:ww:Atree}
  A(\gamma\gamma\to\omega^+\omega^-)_{\rm LO} = %
    2s B(\gamma\gamma\to\omega^+\omega^-)_{\rm LO} = %
    -\frac{1}{t} - \frac{1}{u},
\end{equation}
whereas, at NLO in the counting of figure~\ref{fig:counting},
\begin{subequations}\label{A:scatter:gammagamma:to:ww:Aloop}
\begin{align}    
  A(\gamma\gamma\to\omega^+\omega^-)_{\rm NLO} &= %
    \frac{8(a_1^r-a_2^r+a_3^r)}{v^2} %
    +\frac{2ac_\gamma^r}{v^2} +\frac{a^2-1}{8\pi^2 v^2} \equiv A_C%
\label{A:scatter:gammagamma:to:ww:Aloop:A}  \\
  B(\gamma\gamma\to\omega^+\omega^-)_{\rm NLO} &= 0.
\label{A:scatter:gammagamma:to:ww:Aloop:B}
\end{align}
\end{subequations}
Interestingly, in dimensional regularization all UV divergences cancel after some algebra, so that no renormalization is required and
\begin{subequations}
\begin{align}
  c_\gamma^r &= c_\gamma \\
  a_1^r-a_2^r+ a_3^r  &= a_1-a_2+a_3.\end{align}
\end{subequations}
Thus this particular combination of the chiral parameters $a_1$, $a_2$ and $a_3$ turns out to be finite and the corresponding renormalized one does not depend on any renormalization scale $\mu$.

We assign momenta and polarization vectors $\epsilon_i(\pm)$ to the initial-state vector and final-state scalar bosons as
\begin{equation}
  \{\gamma [\epsilon_1(\pm), k_1], \gamma [\epsilon_2(\pm), k_2]\}\to %
  \{(\omega/h) [p_1], (\omega/h) [p_2]\},
\end{equation}

In the cm frame we may choose the coordinate axes such that
\begin{subequations}\label{A:scatter:gammagamma:def:momenta}
\begin{gather}
  k_1     = (E,0,0,E)    \,\quad %
  k_2     = (E,0,0,-E)   \\
  p_1     = (E,\vec{p})  \,\quad %
  p_2     = (E,-\vec{p}) \,\quad %
  \Delta  = p_1 - p_2    \\
  \vec{p} = (p_x,p_y,p_z) = E (\sin\theta\cos\varphi,\sin\theta\sin\varphi,\cos\theta) .
\end{gather}
\end{subequations}
Because in the cm $\vec{k}_1\parallel \vec{k}_2$, the 4-dimensional polarization states $\epsilon_i(\pm)$ are perpendicular to both, 
$  \epsilon_i(\pm)\cdot k_j = 0\ .
$
This simplifies Eqs.~(\ref{A:scatter:gammagamma:LorentzStructure:1}) and~(\ref{A:scatter:gammagamma:LorentzStructure:2}) to
\begin{equation}\label{A:scatter:gammagamma:LorentzStructure:1:tmp1}
   \left(\epsilon_1^\mu\cdot\epsilon_2^\nu T_{\mu\nu}^{(1)}\right) = %
       \frac{s}{2}\epsilon_1\cdot\epsilon_2\ ,
\end{equation}
and, (since the WBGBs are massless, $(t-u)^2 = s^2\cos^2\theta$), 
\begin{equation}\label{A:scatter:gammagamma:LorentzStructure:2:tmp1}
   \left(\epsilon_1^\mu\cdot\epsilon_2^\nu T_{\mu\nu}^{(2)}\right) = %
       2s(\epsilon_1\cdot\Delta)(\epsilon_2\cdot\Delta) - s^2(\cos\theta)^2 (\epsilon_1\cdot\epsilon_2) \ .
\end{equation}
The product $(\epsilon_1\cdot\epsilon_2)$ appearing satisfies a modified orthogonality relation because the momenta of the two photons are opposite; choosing
\begin{subequations}\label{A:scatter:gammagamma:def:polarization}
\begin{align}
  \epsilon_1(\pm) &= \frac{1}{\sqrt{2}} (0, \mp 1, -i, 0) \\
  \epsilon_2(\pm) &= \frac{1}{\sqrt{2}} (0, \mp 1, i, 0) ,
\end{align}
\end{subequations}
we have $\epsilon_1(+)\cdot\epsilon_2(-)=\epsilon_1(-)\cdot\epsilon_2(+)=0$ and $\epsilon_1(+)\cdot\epsilon_2(+)=\epsilon_1(-)\cdot\epsilon_2(-)=-1$. Thus, the Lorentz structures needed for Eqs.~(\ref{A:scatter:gammagamma:LorentzStructure:1:tmp1}) and~(\ref{A:scatter:gammagamma:LorentzStructure:2:tmp1}) become those shown on table~\ref{A:scatter:gammagamma:comput:LorentzStructure}.
\begin{table}[!h]\centering
\begin{tabular}{l|cccc}
$(\lambda_1\lambda_2)$					& $(++)$ & $(+-)$ & $(-+)$ & $(--)$ \\\hline
$[\epsilon_1^\mu(\lambda_1)\cdot\epsilon_2^\nu (\lambda_2)] T_{\mu\nu}^{(1)}$	& $-s/2$ & $0$    & $0$    & $-s/2$ \\
$[\epsilon_1^\mu(\lambda_1)\cdot\epsilon_2^\nu (\lambda_2)] T_{\mu\nu}^{(2)}$ %
   & $s^2$ & $-s^2(\sin\theta)^2 e^{2i\varphi}$ & $-s^2(\sin\theta)^2 e^{-2i\varphi}$ & $s^2$
\end{tabular}\caption{%
Lorentz structures $\epsilon_1^\mu\cdot\epsilon_2^\nu T_{\mu\nu}^{(1)}$ and $\epsilon_1^\mu\cdot\epsilon_2^\nu T_{\mu\nu}^{(2)}$ (Eqs.~\ref{A:scatter:gammagamma:LorentzStructure:1:tmp1} and~\ref{A:scatter:gammagamma:LorentzStructure:2:tmp1}). %
All are invariant under $\theta\to\pi -\theta$, that is, $t-u$ exchange, a consequence of Bose symmetry that guarantees $\mM(\gamma\gamma\to\omega^+\omega^-)_{\rm LO, NLO}=\mM(\gamma\gamma\to\omega^-\omega^+)_{\rm LO, NLO}$. 
}\label{A:scatter:gammagamma:comput:LorentzStructure}
\end{table}

The structure of table~\ref{A:scatter:gammagamma:comput:LorentzStructure} is remarkable. First, the amplitudes with equal photon helicities, $T^{++}$ and $T^{--}$, come in the combination $-\frac{s}{2} A + s^2 B$. But, due to Eq.~(\ref{A:scatter:gammagamma:to:ww:Atree}), this just cancels the LO contribution, and with it Rutherford's $1/t$ collinear divergence (and the exchange one in $1/u$).  There is no photon-photon annihilation with equal helicity into the Goldstone bosons at LO. Second, the opposite-helicity combinations $T^{+-}$ and $T^{-+}$ are nonvanishing at LO, but again table~\ref{A:scatter:gammagamma:comput:LorentzStructure} assigns a kinematic factor $\sin^2\theta$ that just cancels the angular dependence from the $t$- and $u$-channel $\omega$ exchange diagrams, and thus once more the collinear divergences drop out. Therefore, polar angular integrals may be easily computed and partial wave amplitudes to be introduced in section~\ref{sec:PartialWaves:gammagamma} are well defined for all helicity combinations.

Since we formulate unitarity of the EWSBS in terms of custodial $SU(2)_{L+R}$ isospin partial waves, we use the appropriate Clebsch-Gordan coefficients from Eq.~(\ref{ClebschGordan}) below to obtain the matrix elements that follow. To shorten notation in the next paragraphs, we  use the letters $N$ and $C$ (for ``neutral'' and ``charged'' respectively) to indicate, the $zz$ and $\omega^+\omega^-$ final states, as defined in Eqs.~(\ref{A:scatter:gammagamma:to:zz:Aloop}) and~(\ref{A:scatter:gammagamma:to:ww:Aloop}).
We further shorten $T_I^{\lambda_1\lambda_2}\equiv \Braket{I,0|T|\lambda_1\lambda_2}$; explicitly,
\begin{subequations}\label{T0}
\begin{align}
   T_0^{\lambda_1\lambda_2} &= -\frac{1}{\sqrt{3}}\left(2 T_C^{\lambda_1\lambda_2}+T_N^{\lambda_1\lambda_2}\right) \\ \label{T2}
   T_2^{\lambda_1\lambda_2} &= \frac{2}{\sqrt{6}}\left(T_N^{\lambda_1\lambda_2} - T_C^{\lambda_1\lambda_2}\right) .
\end{align}
\end{subequations}
Taking into account Eq.~(\ref{A:scatter:gammagamma:to:zz:Aloop:A}) through 
(\ref{A:scatter:gammagamma:to:ww:Aloop:B}), we find
\begin{subequations}\label{A:scatter:gammagamma:T_I_lambda1_lambda2}
\begin{align}
   T_0^{++} &= T_0^{--} = \frac{e^2 s}{2\sqrt{3}}\left(2A_C + A_N\right) &
   T_2^{++} &= T_2^{--} = \frac{e^2 s}{\sqrt{6}}\left(A_C - A_N \right) %
 \label{A:scatter:gammagamma:T_I_lambda1_lambda2:1}\\
   T_0^{+-} &= (T_0^{-+})^* = \frac{4e^2}{\sqrt{3}} e^{2i\varphi} &
   T_2^{+-} &= (T_2^{-+})^* = \frac{4e^2}{\sqrt{6}}e^{2i\varphi}\ . %
 \label{A:scatter:gammagamma:T_I_lambda1_lambda2:2}
\end{align}
\end{subequations}

We now turn to the isosinglet scattering amplitude with two Higgses, and obtain
\begin{equation} \label{firstofR}
R(\gamma\gamma\to hh)= -\frac{e^2}{8\pi^2 v^2} (a^2-b)(\epsilon_1\cdot\epsilon_2).
\end{equation}
This is an NLO scattering amplitude as the LO one vanishes. It is proportional to $(a^2-b)$ and thus to the LO crossed-channel $\omega\omega\to hh$. If BSM physics does not couple $hh$ and $\omega\omega$, then, because $hh$ has no charge, it decouples from $\gamma\gamma$ too.

With the polarization vectors of Eq.~(\ref{A:scatter:gammagamma:def:polarization}), Eq.~(\ref{firstofR}) becomes
\begin{equation}
    R(\gamma\gamma\to hh) = \frac{e^2}{8\pi^2 v^2} (a^2-b)\delta_{\lambda_1,\lambda_2} 
\end{equation}
as the final  $\Ket{hh}$ state is an isospin singlet; or explicitly,
\begin{equation}\label{A:scatter:gammagamma:R}
    R_0^{++}=R_0^{--} = \Braket{hh| T(\gamma\gamma\to hh)|++} = \frac{e^2}{8\pi^2 v^2} (a^2-b) .
\end{equation}

\section{Scattering partial waves with $\gamma\gamma$ states}\label{sec:PartialWaves:gammagamma}
In order to unitarize the $\gamma\gamma\to\omega\omega$ scattering amplitudes, we will use the partial wave decomposition
\begin{align}
   P_{IJ}^{\lambda_1\lambda_2} &= \frac{1}{128\pi^2}\sqrt{\frac{4\pi}{2J+1}}\int d\Omega\, T_I^{\lambda_1\lambda_2}(s,\Omega) Y_{J,\Lambda}(\Omega), & \Lambda&=\lambda_1-\lambda_2,
\end{align}
whose inverse is
\begin{equation}\label{chLagr:PartialWavesGeneral:invers}
   T_I^{\lambda_1\lambda_2}(s,\Omega) = 128\pi^2\sum_J\sqrt{\frac{2J+1}{4\pi}}%
       P_{IJ}^{\lambda_1\lambda_2} Y_{J,\Lambda}(\Omega).
\end{equation}
Because of parity conservation\footnote{Note that our $\Ket{\lambda_1\lambda_2}$ state is defined as $\Ket{\lambda_1\lambda_2} = (1/N)\left(\Ket{+k\hat{e}_z,\lambda_1;-k\hat{e}_z,\lambda_2} + \Ket{-k\hat{e}_z,\lambda_2;+k\hat{e}_z,\lambda_1} \right)$. Hence, the parity operator $\mP$ acts according to
$\mP\Ket{\pm\pm}=\Ket{\mp\mp}$, $\mP\Ket{\pm\mp}=\Ket{\pm\mp}$. See Refs.~\cite{Chung:1971ri,JAAguilar,Landau}.}, and for $J=0$, our scattering amplitude only couples to the positive parity state $(\Ket{+-}+\Ket{-+})/\sqrt{2}$. Thus, let us introduce the notation
\begin{equation}
   P_{I0} \equiv \frac{1}{\sqrt{2}}\left(P_{I0}^{++} + P_{I0}^{--}\right) = \sqrt{2}P_{I0}^{++} = \sqrt{2}P_{I0}^{--}.
\end{equation}
For $J=2$, the only non-vanishing contributions come from $P_{I2}^{+-}$ ($\Lambda=+2$) and $P_{I2}^{-+}$ ($\Lambda=-2$). The amplitudes with $\Lambda=0$ vanish (see Eqs.~\ref{A:scatter:gammagamma:T_I_lambda1_lambda2:1} and~\ref{A:scatter:gammagamma:T_I_lambda1_lambda2:2}). Hence, let us define
\begin{equation}
   P_{I2} \equiv P_{I2}^{+-} = P_{I2}^{-2} .
\end{equation}
With these definitions, the lowest nonvanishing-order (denoted with a $(0)$ superindex) $\gamma\gamma$ partial waves are
\begin{subequations}\label{A:scatter:gammagamma:partialWaves:F:e2}
\begin{align}
P_{00}^{(0)} &= \frac{e^2 s}{32\pi\sqrt{6}}(2A_C + A_N) &
P_{02}^{(0)} &= \frac{e^2}{24\pi\sqrt{2}} %
\label{A:scatter:gammagamma:partialWaves:F:e2:1}\\
P_{20}^{(0)} &= \frac{e^2 s}{32\pi\sqrt{3}}(A_C - A_N) &
P_{22}^{(0)} &= \frac{e^2}{48\pi} . %
\label{A:scatter:gammagamma:partialWaves:F:e2:2}
\end{align}
\end{subequations}
Here, the $J=0$ partial waves are NLO while the $J=2$ ones are LO. 

The $hh$ final state is an isospin singlet, and only couples with $J=0$ and positive parity states (see Eq.~\ref{A:scatter:gammagamma:R}). Thus, the corresponding partial waves are
\begin{equation}
   R_I^{(0)} \equiv \frac{1}{\sqrt{2}}\left(R_{I0}^{++}+R_{I0}^{--}\right) = \sqrt{2}R_{I0}^{++}.
\end{equation}
Hence,
\begin{equation}\label{A:scatter:gammagamma:amp:R}
   R_0^{(0)} = \frac{e^2}{128\sqrt{2}\pi^3 v^2}(a^2-b) .
\end{equation}
Finally, let us introduce the fine structure constant $\alpha=e^2/4\pi$ on Eqs.~(\ref{A:scatter:gammagamma:partialWaves:F:e2:1}) and~(\ref{A:scatter:gammagamma:partialWaves:F:e2:2}), so that the $P_{IJ}$ and $R_0$ to NLO turn into
\begin{subequations}\label{A:scatter:gammagamma:partialWaves:F:alpha}
\begin{align}
  P_{00}^{(0)} &= \frac{\alpha s}{8\sqrt{6}}(2A_C + A_N) & %
  P_{02}^{(0)} &= \frac{\alpha}{6\sqrt{2}} %
  \label{A:scatter:gammagamma:partialWaves:F:alpha:1}\\
  P_{20}^{(0)} &= \frac{\alpha s}{8\sqrt{3}}(A_C - A_N) &
  P_{22}^{(0)} &= \frac{\alpha}{12} %
  \label{A:scatter:gammagamma:partialWaves:F:alpha:2}
\end{align}
\begin{equation}\label{A:scatter:gammagamma:partialWaves:F:alpha:3}
R_0^{(0)} = \frac{\alpha}{32\sqrt{2}\pi^2 v^2}(a^2-b) \ .
\end{equation}
\end{subequations}
These last equations are the ones to be used in practice, with $A_c$ and $A_N$ taken from Eqs.~(\ref{A:scatter:gammagamma:to:zz:Aloop}) and~(\ref{A:scatter:gammagamma:to:ww:Aloop}).

\section{Unitarity requires $\omega\omega$ resonances to be visible in $\gamma\gamma$.}
\label{sec:unitarity}
Unitarization of elastic $\omega\omega$ and cross-channel $\omega\omega\to hh$ has been extensively reported in our earlier work~\cite{Delgado:2013loa,Delgado:2015kxa,Delgado:2014dxa} and that of other groups~\cite{Espriu:2013fia,Espriu:2012ih,Espriu:2014jya,Corbett:2015lfa} and will not be repeated here. In this section, we will extend the discussion therein to include the $\gamma\gamma$ channel,
\begin{equation}
 \gamma\gamma  \longleftrightarrow  \{\omega\omega, hh\}\ .  
\end{equation}
The perturbative partial wave amplitudes involving two photons have been given in section~\ref{sec:gammagamma} and their partial-wave projections in section~\ref{sec:PartialWaves:gammagamma} so we have all the necessary ingredients from perturbation theory at hand. As the photon is a spin-1 massless boson, Landau-Yang's theorem forbids the partial wave with $J=1$. Thus, to NLO in the effective theory, the possible angular momenta are $J=0,\,2$. 

The $\omega\omega$ partial waves decouple from the $hh$ channel for $a^2=b$, (see Eq.~\ref{A:scatter:gammagamma:amp:R}). In keeping the more general $a^2\neq b$ situation, the reaction matrix includes an inelastic $\gamma\gamma\to hh$ coupling and is not block diagonal. 

Because the electromagnetic interaction violates custodial isospin conservation (each $\omega$ boson has a different electric charge), the $\gamma\gamma$ state couples to both $I=0$ and $I=2$ (unlike $hh$ which is a singlet $\Ket{0,0} = \Ket{hh}$). Though each channel has its own separate strong dynamics, they both provide probability flow into the $\gamma\gamma$  state as dictated by the corresponding Clebsch-Gordan coefficients. With $\Ket{1,1}\equiv -\omega^+$, $\Ket{1,0}\equiv\omega^0$, $\Ket{1,-1}\equiv\omega^-$, the standard phase conventions~\cite{Rose} and zero total electric charge, $I_3=0$, we have:
\begin{subequations}\label{ClebschGordan}
\begin{align}
   \Ket{0,0} &= -\frac{1}{\sqrt{3}}\left(\Ket{\omega^+\omega^-}+\Ket{\omega^-\omega^+}+\Ket{zz}\right) \\
   \Ket{2,0} &=  \frac{1}{\sqrt{6}}\left(2\Ket{zz}-\Ket{\omega^+\omega^-}-\Ket{\omega^-\omega^+}\right) .
\end{align}
\end{subequations}

\subsection{$\gamma\gamma$ scattering with $hh$ channel decoupled}
\label{chLagr:sec:Unitar:gg:NOhh}

To start, let us decouple the $hh$ channel by setting $a^2=b$ (and other parameters coupling $\omega\omega$ and $hh$ in our earlier work, $d=e=0$). Then, the amplitude matrix is three by three (we specify the $I=0,2$ isospin index to make the three-channel nature of the matrix manifest; for each of them, the angular momentum index $J$ can also take the values 0 or 2). It can be given as
\begin{equation}\label{chLagr:UnitProceduresWW:Fcoupled1}
  F(s) = %
   \begin{pmatrix}
      A_{0J}(s)   & 0          & P_{0J}(s) \\
      0           & A_{2J}(s)  & P_{2J}(s) \\
      P_{0J}(s)   & P_{2J}(s)  & 0
   \end{pmatrix} + \mO(\alpha^2),
\end{equation}
where $A_{IJ}(s)$ are the (isospin conserving) elastic partial waves $\omega\omega\to\omega\omega$ (see ~\cite{Delgado:2013hxa,Delgado:2015kxa} for the exact definition and main properties) and $P_{IJ}(s)$, the partial-wave projected $\gamma\gamma\to\omega\omega$ amplitudes. Note that we consider only the leading order in the electromagnetic coupling $\alpha$. Hence, $\Braket{\gamma\gamma |F^{(0)}|\gamma\gamma} \simeq0$.

The unitarity condition on a coupled-channel problem is 
\begin{equation}\label{unitarity}
\Imag F(s) = F(s) F(s)^\dagger
\end{equation}
on the right cut (RC). Because the interactions among Goldstone bosons grow with $s$ and become strong, a unitarization scheme is mandatory to have a sensible amplitude~\cite{Truong:1988zp}. On the other hand, since  $\alpha$ remains small, it can be considered at leading order so that Eq.~(\ref{unitarity}) will be satisfied to all orders in $s$ but only to LO in $\alpha$, with no appreciable error. Imposing the unitarity condition to Eq.~(\ref{chLagr:UnitProceduresWW:Fcoupled1}) and working to LO in $\alpha$ returns
\begin{subequations}\label{unitarityexpanded}
\begin{align}
   \Imag A_{IJ} &= \lvert A_{IJ}\rvert^2\\
   \Imag P_{IJ} &= P_{IJ}A_{IJ}^*\ .
\end{align}
\end{subequations}
The reader can appreciate that the second of these equations is linear in $P_{I0}$ and does not involve the $\gamma\gamma\to\gamma\gamma$ kernel in the LO approximation in the $\alpha$ expansion. The structure of Eq.~(\ref{unitarityexpanded}) allows to sequentially solve the unitarity equation for elastic $\omega\omega$ scattering and then use it to unitarize the final state $\omega\omega\to \gamma\gamma$ amplitude. According to our Ref.~\cite{Delgado:2015kxa}, the elastic $\omega\omega\to\omega\omega$ amplitude admits a chiral expansion
\begin{equation}
   A(s) = A^{(0)}(s) + A^{(1)}(s) + \mO(s^3),
\end{equation}
where
\begin{subequations}
\begin{align}
   A^{(0)}(s) &= Ks \label{def:A:LO}\\
   A^{(1)}(s) &= \left(B(\mu) + D\log\frac{s}{\mu^2} + E\log\frac{-s}{\mu^2}\right)s^2. \label{def:A:NLO}
\end{align}
\end{subequations}
All the coefficients $K$ (Eq.~\ref{def:A:LO}), $B(\mu)$, $D$ and $E$ (Eq.~\ref{def:A:NLO}) are given in our Ref.~\cite{Delgado:2015kxa} for each partial wave.

The unitarization of the scalar $\omega\omega\to\omega\omega$ ($J=0$) partial-wave is beautifully achieved by the elastic IAM method, constructed from the first two orders of the perturbative expansion $A=A^{(0)}+A^{(1)}+\dots$,
\begin{equation}\label{IAM}
  \tilde{A}(s) = \frac{A^{(0)}(s)}{1-\frac{A^{(1)}(s)}{A^{(0)}(s)}} .
\end{equation}

There is more to this simple equation than meets the eye. It has the correct analytic structure in the complex $s$ plane, allowing for resonances in the second Riemann sheet below the RC, where it satisfies elastic unitarity. At low $\sqrt{s}$ it matches the chiral expansion as can be seen by reexpanding it. And since its derivation follows from a fully prescribed dispersion relation, it can be written down to higher orders (should e.g. the NNLO chiral amplitude become known) without ambiguity.

Turning to  the channel-linking $P$ amplitudes, the second of Eq.~(\ref{unitarityexpanded}) is the statement of Watson's theorem, that sets its phase to that of $\omega\omega$ rescattering. Observing that at low energies, $P\approx P^{(0)}$, its simplest solution with the proper analytical structure is
\begin{equation} \label{Omnes}
   \tilde{P} = \frac{P^{(0)}}{1-\frac{A^{(1)}}{A^{(0)}}} = \frac{P^{(0)}}{A^{(0)}}\tilde{A}\ ,
\end{equation}
which can be obtained from the Omn\`es-Mushkelishvili solution to the dispersion relation or here simply substituted as an ansatz to affirm its validity. Indeed, taking its imaginary part on the RC complies with the second of Eq.~(\ref{unitarityexpanded}),
\begin{equation}\label{chLagr:UnitProceduresWW:Fcoupled1:proof}
   \Imag\tilde{P} = %
     \frac{P^{(0)}}{A^{(0)}}\Imag\tilde{A} = %
     \frac{P^{(0)}}{A^{(0)}}\lvert\tilde{A}\rvert^2 = %
     \tilde{P}\tilde{A}^*,
\end{equation}
where Eq.~(\ref{IAM}) was substituted in the last step. Thus, our unitarized $\gamma\gamma\to\omega\omega$ matrix element will be
\begin{equation}\label{chLagr:UnitProceduresWW:Fcoupled1:fin}
  \tilde{P}_{I0} = \frac{P_{I0}^{(0)}}{1-\frac{A_{I0}^{(1)}}{A_{I0}^{(0)}}},\quad I=0,2 .
\end{equation}
The computation of the $P_{I0}^{(0)}$ partial waves for the $\gamma\gamma\to\omega\omega$ can be found in section~\ref{sec:PartialWaves:gammagamma}.

Now we deal with the tensor $J=2$ channel: note here that the $P_{I2}$ are constant. Also, we have a vanishing LO elastic $\omega\omega$ scattering amplitude $A_{I2}=K_{I2}s$ because $K_{I2}=0$. And due to $A^{(0)}=0$, the IAM unitarization method in Eq.~(\ref{IAM}) cannot be applied. Hence, the N/D method will be used here. In the scalar channel we know that both methods (as well as others), provide very similar solutions to the IAM (see Ref.~\cite{Delgado:2015kxa}). A quick, algebraic way to construct an approximation to the $N/D$ system of dispersion relations that satisfies elastic unitarity for all $s$ and has the right analytic properties, having only at hand one order of perturbation theory (here, the NLO) is
\begin{equation}\label{unitar:ND:elastic}
\tilde{A} = A^{\rm N/D} = \frac{A_L(s)}{1+\frac{1}{2}g(s)A_L(-s)},
\end{equation}
where
\begin{subequations}
\begin{align}
   g(s)   &= \frac{1}{\pi}\left(\frac{B(\mu)}{D}+\log\frac{-s}{\mu^2}\right)       \label{unitar:ND:elastic:g}\\
   A_L(s) &= \left(\frac{B(\mu)}{D} + \log\frac{s}{\mu^2}\right) D s^2 = \pi g(-s)Ds^2  . \label{unitar:ND:elastic:AL}
\end{align}
\end{subequations}

The $B$ and $D$ which appear in Eq.~(\ref{unitar:ND:elastic:AL}) is the same that those in Eq.~(\ref{def:A:NLO}). Note that, by means of perturbative unitarity, $K=0\implies E=0$, thus simplifying the full N/D expression of Ref.~\cite{Delgado:2015kxa}.

Once the $J=2$ elastic $\omega\omega$ waves have been unitarized, it is easy to satisfy
the second of Eq.~(\ref{unitarityexpanded}) by 
\begin{equation}\label{chLagr:UnitProceduresWW:Fcoupled3:fin}
   \tilde{P}_{I2} = \frac{P_{I2}^{(0)}}{A_{{\rm L},I2}}A_{I2}^{\rm N/D},%
   \quad I=0,2 .
\end{equation}
For $J=0$ we need to use the full expressions of~\cite{Delgado:2015kxa}.

\subsection{Coupled $\gamma\gamma \longleftrightarrow (\omega\omega, hh)  $ scattering}
We now proceed to an analysis of the coupled $\omega\omega$ (that is, $W_LW_L$ and $Z_LZ_L$ as per the ET) and $hh$ channels feeding the $\gamma\gamma$ state. Because the electromagnetic interaction violates isospin conservation, the reaction matrix must include both $I=0,2$ subchannels of the $\omega\omega$ system, and is thus of dimension four.
Assuming weak isospin conservation in the Goldstone dynamics, which puts zeroes in row three and column three, and to order $\alpha$, which makes the $(4,4)$ element vanish, it is
\begin{equation}\label{chLagr:UnitProceduresWW:Fcoupled2}
  F = %
   \begin{pmatrix}
      A_{0J}   & M_J     & 0       & P_{0J} \\
      M_J      & T_J     & 0       & R_J    \\
      0        & 0       & A_{2J}  & P_{2J} \\
      P_{0J}   & R_J     & P_{2J}  & 0
   \end{pmatrix} + \mO(\alpha^2)\ .
\end{equation}
Here again, $A_{IJ}(s)$ are the partial waves $\omega\omega\to\omega\omega$; $M_J(s)$, the $\omega\omega\to hh$ partial wave; $T_J(s)$, the elastic $hh\to hh$ one; $P_{IJ}(s)$, the $\gamma\gamma\to\omega\omega$ ones (Eqs.~\ref{A:scatter:gammagamma:partialWaves:F:alpha:1} and~\ref{A:scatter:gammagamma:partialWaves:F:alpha:2}) that we newly incorporate in the unitarization in this work; and $R_J(s)$, the $\gamma\gamma\to hh$ (Eq.~\ref{A:scatter:gammagamma:partialWaves:F:alpha:3}).

On the RC, the unitarity  relations in Eq.~(\ref{unitarity}), perturbative in $\alpha$, can be split into three blocks, the Eqs.~(\ref{chLagr:sec:Unitar:gg:YEShh:J0:new_unit_relat}), (\ref{chLagr:sec:Unitar:gg:YEShh:J0:new_unit_relat2}) and~(\ref{chLagr:sec:Unitar:gg:YEShh:J0:new_unit_relat3}) that follow; first, those for the imaginary parts of the elastic amplitudes,
\begin{subequations}\label{chLagr:sec:Unitar:gg:YEShh:J0:new_unit_relat}
\begin{align}
   \Imag A_{0J} &=  \lvert A_{0J}\rvert^2 + \lvert M_J\rvert^2 \\
   \Imag A_{2J} &=  \lvert A_{2J}\rvert^2 \\
   \Imag M_J    &=  A_{0J}M_J^* + M_J T_J^* \\
   \Imag T_{J\ }&=  \lvert M_J\rvert^2 + \lvert T_J\rvert^2 ,
\end{align}
\end{subequations} 
which need to be solved as a coupled-channel problem with all channels being presumably strong. Only then is the solution fed to the second block for the $\gamma\gamma$ couplings, as we work to LO in $\alpha$,
\begin{subequations}\label{chLagr:sec:Unitar:gg:YEShh:J0:new_unit_relat2}
\begin{align}
   \Imag P_{0J} &= P_{0J}A_{0J}^* + R_J M_J^* \\
   \Imag R_J &= P_{0J}M_J^* + R_J T_J^* .
\end{align}
\end{subequations}  
Finally, the isotensor block decouples from the isoscalar ones and becomes
\begin{subequations}\label{chLagr:sec:Unitar:gg:YEShh:J0:new_unit_relat3}
\begin{align}
   \Imag A_{2J} &= \lvert A_{2J}\rvert^2 \\
   \Imag P_{2J} &= P_{2J}A_{2J}^* ,
\end{align}
\end{subequations}
which is identical to Eq.~(\ref{unitarityexpanded}) and can be solved with the methods of subsection~\ref{chLagr:sec:Unitar:gg:NOhh}, so we concentrate in what follows only in the first two blocks corresponding to isospin 0.

The previous discussion of section~\ref{chLagr:sec:Unitar:gg:NOhh} can be mimicked easily also for $I=0$ by writing down first a reaction submatrix for the strongly interacting subchannel ${}_sF(\omega\omega,hh\to \omega\omega,hh)$,
\begin{equation}\label{chLagr:UnitProceduresWW:Fcoupled2:tmp0}
   {}_sF =  %
     \begin{pmatrix}
       A_{00} & M_0 \\
       M_0    & T_0
     \end{pmatrix} \equiv %
     \begin{pmatrix}
       A & M \\
       M & T
     \end{pmatrix},
\end{equation}
definition that can analogously be adopted for the matrices in the low-$s$ chiral expansion, $_sF^{(0)}$ and $_sF^{(1)}$.

We now need to distinguish the cases $J=0$ and $J=2$, and handle the first right away.
The matricial generalization of the IAM method in Eq.~(\ref{IAM}) yields a unitary $_s\tilde{F}$, from knowledge of the first two terms in the chiral expansion, 
\begin{equation}
{}_s\tilde{F}={}_sF^{(0)}({}_sF^{(0)}-{}_sF^{(1)})^{-1} {}_sF^{(0)}\ .
\end{equation}
The matrix elements of the IAM subreaction matrix can be likewise signaled with a tilde,
\begin{equation}\label{IAMpieces}
   _s\tilde{F} = %
     \begin{pmatrix}
       \tilde{A} & \tilde{M} \\
       \tilde{M} & \tilde{T}
     \end{pmatrix}\ .
\end{equation}
This IAM approximation to the exact $_sF$ in Eq.~(\ref{chLagr:UnitProceduresWW:Fcoupled2:tmp0}) has all relevant properties: unitarity in the RC, {\it i.e.} $\Imag _s\tilde{F} = _s\tilde{F}\cdot _s\tilde{F}^\dagger = _s\tilde{F}^\dagger\cdot _s\tilde{F}$, analyticity, and matching to the chiral expansion at NLO. 

If we also shorten notation $(P,R)\equiv (P_{00}, R_0)$, then from Eqs.~\ref{chLagr:sec:Unitar:gg:YEShh:J0:new_unit_relat},
\begin{equation}
   \Imag\begin{pmatrix} P \\ R\end{pmatrix} = {}_sF^*\cdot \begin{pmatrix} P \\ R\end{pmatrix}\ .
\end{equation}
This is solved by the  unitarized amplitude generalizing Eq.~(\ref{Omnes}),
\begin{equation}\label{chLagr:UnitProceduresWW:Fcoupled2:raw}
   \begin{pmatrix}\tilde{P}\\\tilde{R}\end{pmatrix} \equiv %
      {}_s\tilde{F}({}_sF^{(0)})^{-1}\begin{pmatrix}P^{(0)}\\R^{(0)}\end{pmatrix}.
\end{equation}
Using the chiral expansion ${}_s\tilde{F} = {}_sF^{(0)} + {}_sF^{(1)} + \dots$, Eq.~\ref{chLagr:UnitProceduresWW:Fcoupled2:raw} turns into
\begin{multline}\label{chLagr:UnitProceduresWW:Fcoupled2:dev2}
   \begin{pmatrix}\tilde{P}\\\tilde{R}\end{pmatrix} \equiv %
      \tilde{F}(F^{(0)})^{-1}\begin{pmatrix}P^{(0)}\\R^{(0)}\end{pmatrix} = %
      (F^{(0)}+F^{(1)}+\dots)(F^{(0)})^{-1}\begin{pmatrix}P^{(0)}\\R^{(0)}\end{pmatrix} \\= %
      \begin{pmatrix} P^{(0)}\\ R^{(0)}\end{pmatrix} + F^{(1)}(F^{(0)})^{-1}%
                                \begin{pmatrix} P^{(0)}\\ R^{(0)}\end{pmatrix}+\dots.
\end{multline}

Note that Eqs.~(\ref{A:scatter:gammagamma:partialWaves:F:alpha:1}), 
(\ref{A:scatter:gammagamma:partialWaves:F:alpha:2}) and~(\ref{A:scatter:gammagamma:partialWaves:F:alpha:3})
explicitly show the perturbative order
\begin{equation}
   \begin{pmatrix}P^{(0)}\\R^{(0)}\end{pmatrix}%
      \sim\begin{pmatrix}\mO\left(\frac{s}{v^2}\right)+ \mO(\alpha)\\\mO(\alpha)\end{pmatrix},
\end{equation}
which excludes intermediate 2-photon states. Higher order contributions in $s$ coming from the WBGBs and $h$ rescatterings are taken into account in the IAM. For example, in expanding to one more order in Eq.~(\ref{chLagr:UnitProceduresWW:Fcoupled2:dev2}) we find
\begin{equation}
    F^{(1)}(F^{(0)})^{-1}\begin{pmatrix} P^{(0)}\\ R^{(0)}\end{pmatrix}%
      \sim\begin{pmatrix}\mO\left(\frac{s^2}{v^4}\right)+ \mO(\alpha)\\\mO\left(\frac{s}{v^2}\right)+\mO(\alpha)\end{pmatrix},
\end{equation}
as required. Eq.~(\ref{chLagr:UnitProceduresWW:Fcoupled2:raw}) may be explictly spelled out as
\begin{subequations}\label{chLagr:UnitProceduresWW:Fcoupled2:fin}
\begin{align}
   \tilde{P} &= P^{(0)}\frac{\tilde{A} T^{(0)} - \tilde{M}M^{(0)}}{A^{(0)}T^{(0)}-(M^{(0)})^2} %
               +R^{(0)}\frac{-\tilde{A}M^{(0)} + \tilde{M}A^{(0)}}{A^{(0)}T^{(0)}-(M^{(0)})^2} \\
   \tilde{R} &= P^{(0)}\frac{\tilde{M} T^{(0)} - \tilde{T}M^{(0)}}{A^{(0)}T^{(0)}-(M^{(0)})^2} %
               +R^{(0)}\frac{\tilde{T}A^{(0)}  - \tilde{M}M^{(0)}}{A^{(0)}T^{(0)}-(M^{(0)})^2},
\end{align}
\end{subequations}
in terms of the $IJ=00$ IAM $\tilde{A}$, $\tilde{M}$ and $\tilde{T}$ partial wave amplitudes of Eq.~(\ref{IAMpieces}).

Finally, for $J=2$ we find once more that the IAM method cannot be constructed without knowledge of the NNLO (in $s$) amplitude, so that the N/D coupled channel method is used instead for the unitarization of the WBGBs and $h$ scattering matrix elements ($A_{I2}$, $M_2$ and $T_2$). The matricial N/D formula, analogous to the elastic case of Eq.~(\ref{unitar:ND:elastic}), is
\begin{equation}\label{NDpieces}
   _s\tilde{F} = %
      \left[1 +\frac{1}{2}G(s)F_L(-s)\right]^{-1}F_L(s) ,
\end{equation}
where
\begin{subequations}
\begin{align}
   G(s)   &= \frac{1}{\pi}\left[B(\mu)D^{-1}+\log\frac{-s}{\mu^2}\right]		\\
   F_L(s) &= \left[B(\mu)D^{-1}+\log\frac{s}{\mu^2} \right]Ds^2 = \pi G(-s)Ds^2	
  \end{align}
\end{subequations}
are the matricial versions of Eqs.~(\ref{unitar:ND:elastic:g}) and following. Note that, although we are in a coupled channel case in the sense that $\omega\omega\to hh\to\omega\omega$ rescattering takes place, $hh$ states do not couple with $\gamma\gamma$ for $J=2$ [see Eq.~(\ref{A:scatter:gammagamma:R})]. Thus, we need the matricial N/D method of Eq.~(\ref{NDpieces}) for unitarizing the $\omega\omega\to\omega\omega$ partial waves, but the coupling with $t\bar t$ states can be computed by using the (scalar) Eq.~\ref{chLagr:UnitProceduresWW:Fcoupled3:fin}.

Finally, for the purpose of cross-checking the IAM in the $J=0$ case, the $P_{02}$ matrix elements can be estimated via the coupled-channel N/D by a matrix analogous of Eq.~(\ref{chLagr:UnitProceduresWW:Fcoupled3:fin}),
\begin{equation}\label{chLagr:UnitProceduresWW:Fcoupled4:fin}
   \tilde{P}_{I2} = {}_s\tilde{F}_{I2} \left( F_{{\rm L},I2}\right)^{-1} P_{I2}^{(0)}\ , %
   \quad I=0,2 \ ,
\end{equation}
with $\tilde{P}_{I2}$ a column vector of two components $\tilde{P}$, $\tilde{R}$.


\section{Some numerical examples}\label{sec:numerics}

\begin{figure}[h]
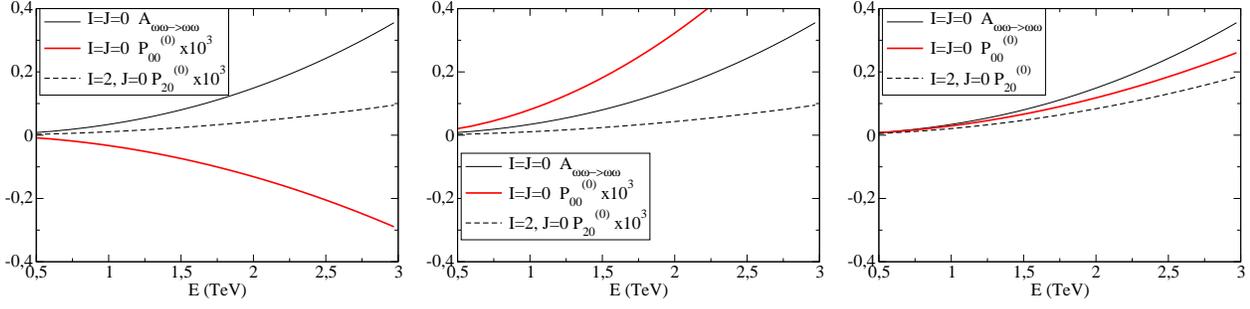

\includegraphics*[width=5.3cm]{FIGS.DIR/PerturbativePs.eps}\hspace{0.2cm}
\includegraphics*[width=5.3cm]{FIGS.DIR/PerturbativePsCgamma.eps}\hspace{0.2cm}
\includegraphics*[width=5.3cm]{FIGS.DIR/PerturbativePsa1a2a3.eps}
\caption{\label{fig:pertamps}
Perturbative $\omega\omega\to \gamma\gamma$ amplitudes driven by the elastic $A(\omega\omega\to\omega\omega)$ shown as the thin line and equal on all plots for reference (corresponding to $a=0.95$). Left: the coupling to the $\gamma\gamma$ sector is produced only by $\alpha$. Center: setting also $c_\gamma=0.5/(16\pi^2)$ but $a_1=a_2=a_3=0$. Right: $c_\gamma=0=a_1$, $a_3-a_2=0.3$. Note that in the first two plots the $P$ amplitudes have been multiplied by a factor $10^3$ for visibility; not so in the third plot.}
\end{figure}

We start by commenting the perturbative partial-wave amplitudes very briefly. Referring to Eq.~(\ref{A:scatter:gammagamma:partialWaves:F:alpha}), we see that the NLO perturbative amplitudes $P^{(0)}_{02}$, $P^{(0)}_{22}$ and $R^{(0)}_0$ are all constant, so we do not plot them. The two $P^{(0)}$ amplitudes coming from the isoscalar $\omega\omega$ state,
quadratic in energy, are shown in figure~\ref{fig:pertamps}. Therein and in what follows we have taken $\alpha_{\rm EM}(Q^2=0)=\frac{1}{137}$ as the emitted photons are real. From the parameters of the EWSBS, we have taken all NLO coefficients to zero, and $b=a^2$, so that the only slight separation from the SM is driven by $a=0.95$; the further parameters of the photon sector are indicated in the figure.

All the amplitudes shown in the figure display the expected quadratic growth with energy (linearity in $s$). Eventually they must violate the unitarity bounds, for example by $\lvert A_{00}\rvert > 1$ which occurs already below $3\,{\rm TeV}$ if we increase $1-a$ or other parameters of the HEFT.

The first two plots show $P$ amplitudes that are much smaller than the elastic $A$ amplitude, as demanded by the smallness of $\alpha$. On the contrary, the third plot exposes values of $P$ of the same order of those of $A$. This means that the {\it a priori} counting of subsection~\ref{subsec:couplegammagamma} fails for the value of $(a_1-a_2+a_3)$ chosen around 0.3; this maximum value allowed by previous constraints is too large in comparison to the ``natural'' values of the $a_i$, of order $10^{-3}$. We do not employ such large values again later.

\begin{figure}[h] 
\centerline{ \includegraphics*[width=8cm]{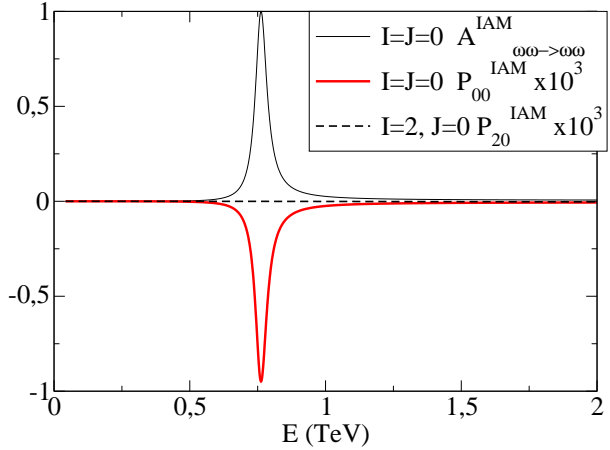} }
\caption{\label{fig:narrowres}With $a=0.81$, $b=a^2$, $a_5(\mu=0.75{\rm TeV})=0.0023$ and other NLO parameters at that scale set to zero, the IAM accommodates a narrow resonance in the scalar channel. We show the imaginary part of the elastic amplitude and the imaginary part of the scalar photon amplitudes (multiplied by $10^3$ as in figure~\ref{fig:pertamps}).}
\end{figure}

The approach that we have developed can be used to describe resonances that could be found in experimental data from the LHC and relate them between different channels. 
Figure~\ref{fig:narrowres} shows a narrow resonance, with $\Gamma/M\sim 0.06$. This is useful to make contact with the large body of theoretical work following the $\gamma\gamma$ statistical fluctuation in the CMS and ATLAS data (we next proceed to more phenomenologically viable resonances). 

Although the electromagnetic interactions do not conserve weak isospin, our choice of $P_{00}$ and $P_{20}$ amplitudes makes it that only the first is fed by a scalar resonance in the $\omega\omega$ channel, as is patent in the figure.
 
The signs of the imaginary parts of $P_{00}$ and $A_{00}$ are seen to be opposite. This is a consequence of $a^2-1<0$ for the choice $a=0.81$ and Eq.~(\ref{A:scatter:gammagamma:to:zz:Aloop:A}), (\ref{A:scatter:gammagamma:to:ww:Aloop:A}) and~(\ref{Omnes}).

\begin{figure}[h]
\centerline{ \includegraphics*[width=8cm]{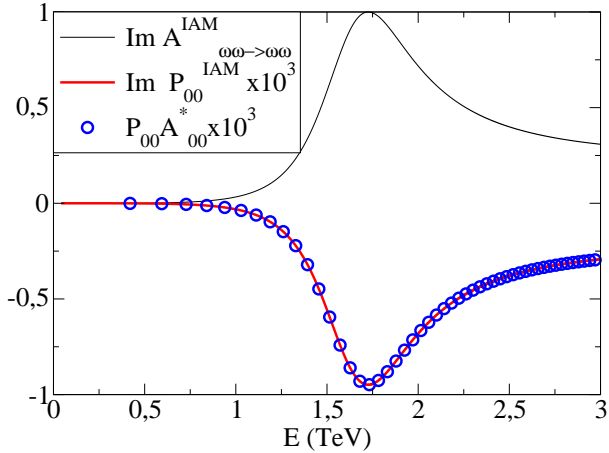}}
\caption{\label{fig:broadelasticres}
Example broad resonance generated by elastic $\omega\omega$ scattering that illustrates the unitarity of $P_{00}$ in the absence of coupling to the $hh$ channel. The parameters employed are $a=0.81$, $b=a^2$, $a_4(\mu=3{\rm TeV})=4\times 10^{-4}$. (All other NLO parameters vanish at that scale.)
}
\end{figure}
Next, we provide an example of a typical broad resonance in figure~\ref{fig:broadelasticres}.

Once more, the only nonvanishing parameter for the two-photon sector is $\alpha=1/137$. The imaginary part of $P_{00}$ (note it has again been multiplied by $10^3$) presents a clear resonating shape driven by that of $A$. We also show how well the unitarity relation of Eq.~(\ref{unitarityexpanded}) is satisfied by our numerical program: the IAM is indeed up to the task, with unitarity satisfied exactly in $s$ and to first order in $\alpha$.

In figure~\ref{fig:depCgamma} we show the dependence of $P_{00}$ on the parameter $c_\gamma$ within its allowed $2\sigma$ band.
\begin{figure}[h]
\centerline{\includegraphics*[width=8cm]{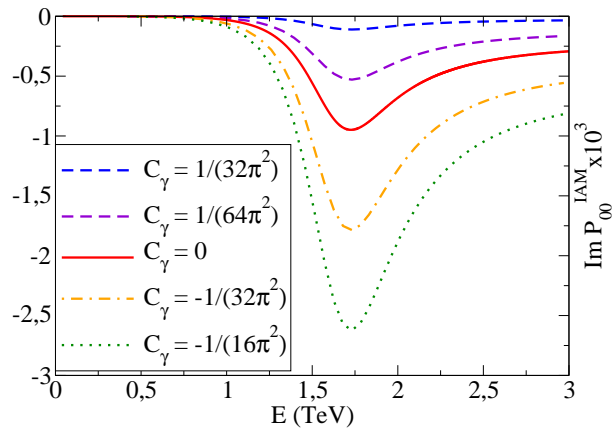}}
\caption{\label{fig:depCgamma}Dependence of $P_{00}$ on $c_\gamma$.}
\end{figure}
The solid line corresponds to $c_\gamma=0$. Positive values thereof diminish the intensity of $P_{00}$, negative values increase it. While the line shape of the resonance is also affected by the value of $c_\gamma$, the position of the maximum (controlled by the IAM $\omega\omega$ amplitude) is not.

In turn, figure~\ref{fig:depa1a2a3} shows the dependence on values of the $(a_1-a_2+a_3)$ parameter combination that are way smaller (of order $\sim 10^{-3}$) than the maximum allowed by the $2\sigma$ bounds (as argued above, values of order $0.1$ are unnaturally large and overturn the counting that we follow).
\begin{figure}[h]
\centering
\includegraphics*[width=8cm]{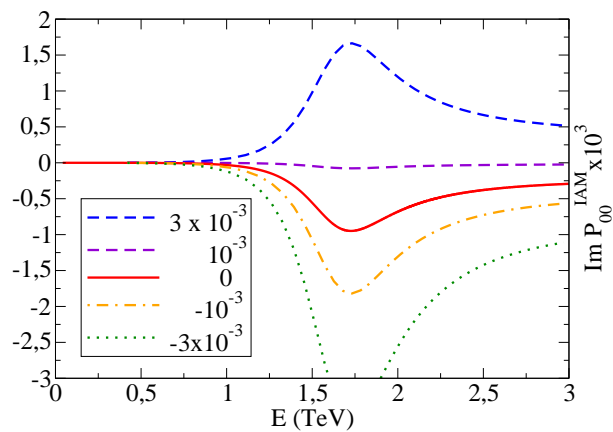}
\caption{\label{fig:depa1a2a3} Dependence of the $P_{00}$ amplitude from figure~\ref{fig:broadelasticres} on $(a_1-a_2+a_3)$ (values thereof are given in the legend).}
\end{figure}
Once more, while the position of the pole is the same for all curves, the line shape and especially the total normalization of the curve do depend on this $a_i$ parameter combination.

We now move on to a coupled channel example, illustrated in figure~\ref{fig:coupledchannel}.
\begin{figure}[h]
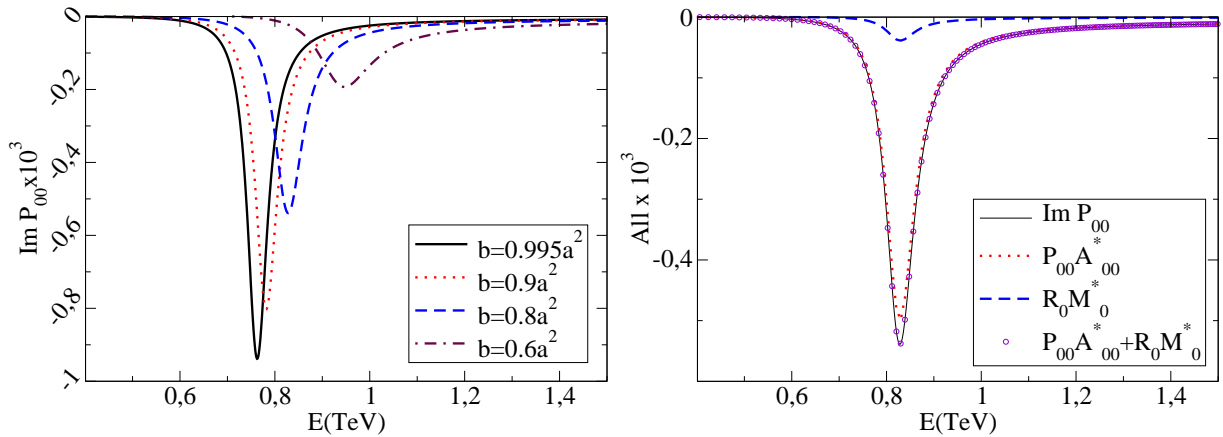

\centerline{
\includegraphics*[width=8cm]{FIGS.DIR/CoupledChannels1.eps}
\includegraphics*[width=8cm]{FIGS.DIR/CoupledChannels2.eps}
}
\caption{\label{fig:coupledchannel}
Left: the resonance of figure~\ref{fig:narrowres} is coupled to the $hh$ with the strength $(b-a^2)$ indicated in the legend for each line, the other parameters remaining the same as earlier.
Right: test of the unitarity relation in Eq.~(\ref{chLagr:sec:Unitar:gg:YEShh:J0:new_unit_relat2}) for the case $b=0.8a^2$.
}
\end{figure}
As $a^2-b$ increases, the $hh$ channel is coupled with larger LO strength. This makes the resonance of figure~\ref{fig:narrowres}, that we take to exemplify, broader and somewhat less intense in the $\gamma\gamma$ channel, resembling more and more the coupled-channel resonance described in~\cite{Delgado:2014dxa}.

The right panel of the same figure then serves to demonstrate unitarity as per the first of Eq.~(\ref{chLagr:sec:Unitar:gg:YEShh:J0:new_unit_relat2}). We see in the plot how elastic unitarity according to Eq.~(\ref{unitarityexpanded}) fails, and how the addition of the $R_0 M^*_0$ component is precisely what achieves coupled-channel unitarity.

Figure~\ref{fig:comparemethods} permits a comparison of the IAM and N/D methods in a case where both are applicable, the scalar-isoscalar channel. We have taken $b$ very close $a^2$ to avoid the coupled-channel complication here (already shown to work in figure~\ref{fig:coupledchannel}).
\begin{figure}[h]
\centerline{\includegraphics[width=8cm]{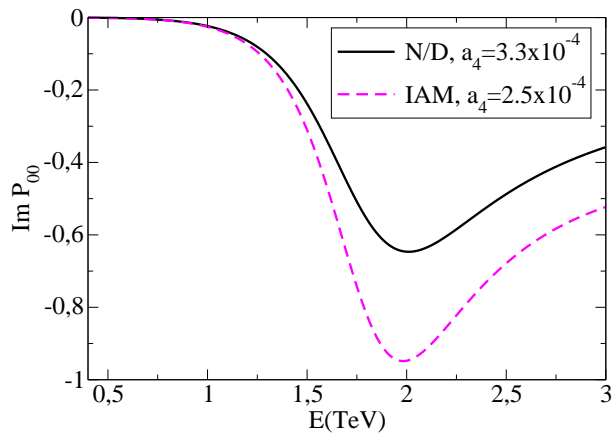}}
\caption{\label{fig:comparemethods} Comparison of the IAM and N/D methods. The LO parameters are $a=0.81$, $b\simeq a^2$; the only nonvanishing NLO parameter at $\mu=3\,{\rm TeV}$ is $a_4$ as indicated for each of the two lines.}
\end{figure}
With both methods we have varied $a_4$ until a scalar resonance appears at $2\,{\rm TeV}$. The necessary value of this parameter is somewhat different, by 25\%. The width and the minimum value of the $P_{00}$ amplitude are not identical either (which could be perhaps arranged by varying some of the other NLO parameters, but we do not see the need at this stage). In conclusion, while both methods give qualitative similar results, their comparison gives us a warning that there is a systematic error in the choice of unitarization scheme of order one part in four. 

\begin{figure}[h]
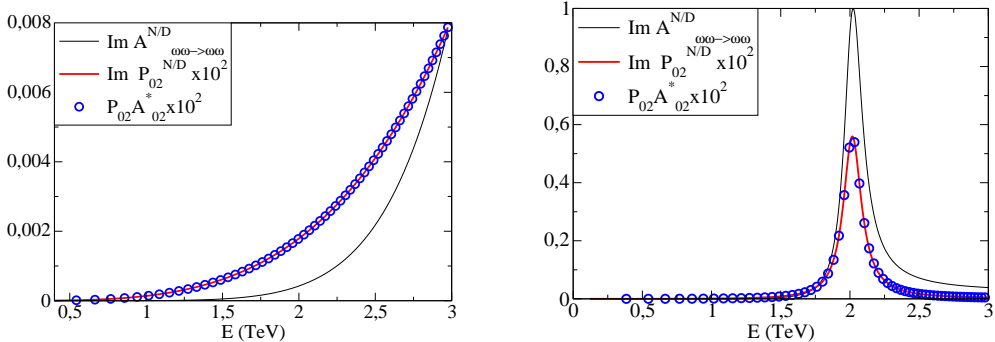

\centerline{
\includegraphics*[width=6cm]{FIGS.DIR/NsobreDPstensor.eps}\hspace{1cm}
\includegraphics*[width=6cm]{FIGS.DIR/NsobreDPstensorRes.eps}
}
\caption{\label{fig:tensor} $P_{02}$ isoscalar, tensor amplitude in the N/D method. Left: same parameters as in figure~\ref{fig:broadelasticres}. Right: we increase $a_4=3\times 10^{-3} $ to induce a tensor resonance in the elastic $\omega\omega$ amplitude that is neatly reproduced in the photon-photon amplitude. In both cases we test unitarity.}
\end{figure}

Finally, in figure~\ref{fig:tensor} we show $P_{02}$, the $\gamma\gamma$ amplitude with the initial $\omega\omega$ in the isoscalar-tensor channel. The left plot is dedicated to show a nonresonant tensor amplitude; for ease of comparison, we employ the same parameters that produced a scalar resonance in figure~\ref{fig:broadelasticres}. An interesting remark is that at relatively low energies the ratio  between the photon production amplitude and the elastic $\omega\omega$ one, $P_{02}/A_{02}$, is much more sizeable than its scalar counterpart  $P_{00}/A_{00}$. This comes about because in Eq.~\ref{A:scatter:gammagamma:partialWaves:F:alpha:1} there is an electromagnetic $\alpha$ coupling as in all the photon amplitudes, but it is constant ($s$ independent) as opposed to $A_{02}$ that has an Adler zero at $s=0$. Therefore both amplitudes can be shown in the same plot by enhancing the photon one only a factor $10^2$, whereas in the scalar case we have been employing $10^3$. 

Many more example calculations are interesting, but we content ourselves with these examples until experimental data shows whether there is merit in pursuing further computations, and specifically which ones. 

\section{Conclusions}\label{sec:conclusions}

In this article we have coupled the EWSBS described with HEFT (for $E<4\pi v\sim 3\,{\rm TeV}$) and the equivalence theorem (for energies $E>M_h,\ M_W$), in the regime of unitarity saturation and resonances, to the two-photon channel which is a promising detection avenue for new physics.

We have developed the necessary unitarization formalism with two different, well explored methods (IAM and N/D), that are equivalent (up to NNLO) to the NLO perturbative amplitudes of~\cite{Delgado:2014jda} at low energies but that, unlike those of the HEFT, can be employed to describe any resonances of the EWSBS.

For example, in figures~\ref{fig:narrowres} and~\ref{fig:broadelasticres} we have shown that both a light, narrow, and a heavier, broader resonance feeding the $\gamma\gamma$ spectrum can be parametrized in this approach, in terms of $a$, $a_4$ and $a_5$ that control the EWSBS. What the production cross section is for those particular resonances is work of phenomenological interest that we postpone to imminent work within an expanded collaboration.

Our formalism assumes that the symmetry-breaking dynamics in the $W$, $Z$ and $h$ sector is stronger than their electromagnetic coupling to $\gamma$s. Nevertheless we have also considered the NLO counterterms that arise in coupling $\gamma\gamma$ to the EWSBS. As long as their values remain ``natural'', our counting in figure~\ref{fig:counting} suggests that perturbation theory is valid in coupling $\gamma\gamma$, and that the resonating $\omega\omega$ (and/or $hh$) amplitudes can be separately computed first. Our theory satisfies Watson's final state theorem in that the phases of the photon-photon production amplitude coincide with those of the elastic EWSBS amplitudes.

A technical challenge that we have overcome is that of projecting the (earlier known) Feynman amplitudes into $\gamma\gamma$-helicity amplitudes of definite total angular momentum $J$ and stemming from $\omega\omega$ states of definite custodial isospin $I$. 
This was necessary and convenient as any resonance or new particle produced in the custodially-invariant EWSBS sector will have specific $I$, $J$ but the detection in the $\gamma\gamma$ channel loses memory of $I$; a complete set of observables however includes the photon helicities $\lambda_1$ and $\lambda_2$ (though for many cross-section calculations one may sum them).

Even in the absence of new resonances, the set of projected  amplitudes that we have provided can be useful to parametrize separations from the SM in a relatively low-energy regime below $3\,{\rm TeV}$, where the partial wave series converges quickly. 

We are currently collaborating with other authors in the preparation of a document with simple estimates for collider cross-sections of typical resonances as seen in the two-photon channel.

Finally, another natural final-state channel that couples with sufficient intensity to the EWSBS, and may serve as LHC probe thereof, is the $t\bar t$ one. We are separately exploring it with the same methods and have recently shown that within HEFT this channel coupling admits a perturbative expansion in powers of $M_t/\sqrt{s}$, obtaining the unitarized amplitudes needed for its description in the resonance region~\cite{Castillo:2016erh}. The calculation follows lines analogous to those here presented.

\section*{Acknowledgements}
We thank useful conversation and suggestions from J.J.~Sanz-Cillero, D.~Espriu, and M.J.~Herrero. Work supported by Spanish grants MINECO:FPA2011-27853-C02-01, MINECO:FPA2014-53375-C2-1-P and by  BES-2012-056054 (RLD).



%

\end{document}